\documentclass[twocolumn,prb,superscriptaddress,amsmath,amssymb,showkeys,showpacs,nofootinbib]{revtex4-2}
\usepackage{natbib}
\usepackage{graphicx}
\usepackage{dcolumn}
\usepackage{bm}
\usepackage[utf8]{inputenc}
\usepackage[T1]{fontenc}
\usepackage{mathptmx}
\usepackage{etoolbox}
\usepackage{bigints}

\usepackage{easyReview}
\usepackage{microtype}
\usepackage{upgreek}
\usepackage{booktabs}
\usepackage{multirow}
\usepackage{tabularx}
\usepackage{xcolor}
\definecolor{link}{RGB}{57,106,177}
\definecolor{blue}{RGB}{0,0,0}
\usepackage{hyperref}
\hypersetup{
	colorlinks=true,
	linkcolor=link,
	citecolor=link,
	urlcolor=link
}
\usepackage{cleveref}
\usepackage{wasysym}
\newcommand{\ped}[1]{\ensuremath{_{\rm #1}}}
\newcommand{\apex}[1]{\ensuremath{^{\rm #1}}}

\begin{document}

\title{A model for critical current effects in point-contact Andreev-reflection spectroscopy}

\author{Dario Daghero}
\affiliation{\mbox{Department of Applied Science and Technology, Politecnico di Torino, I-10129 Torino, Italy}}
\author{Erik Piatti}
\affiliation{\mbox{Department of Applied Science and Technology, Politecnico di Torino, I-10129 Torino, Italy}}
\author{{\color{blue}Nikolai D. Zhigadlo}}
\affiliation{\mbox{{\color{blue}CrystMat Company, CH-8037 Zurich, Switzerland}}}
\author{Renato S. Gonnelli}
\email{renato.gonnelli@polito.it}
\affiliation{\mbox{Department of Applied Science and Technology, Politecnico di Torino, I-10129 Torino, Italy}}

\begin{abstract}
It is well known that point-contact Andreev reflection spectroscopy provides reliable measurements of the energy gap(s) in a superconductor when the contact is in the ballistic or 
{\color{blue}diffusive} regime. However, especially when the mean free path of the material under study is small, obtaining ballistic contacts can be a major challenge. One of the signatures of a Maxwell contribution to the contact resistance {\color{blue}$R$} is the presence of "dips" in the differential conductance, associated to the sudden appearance of a Maxwell term, in turn due to the attainment of the critical current of the material in the contact region. Here we show that, using a proper model for the $R(I)$ of the material under study, it is possible to fit the experimental curves (without the need of normalization) obtaining the correct values of the gap amplitudes even in the presence of such dips, as well as the temperature dependence of the critical current in the contact. We present a test of the procedure in the case of Andreev-reflection spectra in {\color{blue}{Mg$_{0.85}$Al$_{0.15}$B$_2$ }} single crystals.\\\\
Cite this article as: D. Daghero, E. Piatti, N. D. Zhigadlo, and R. S. Gonnelli. \textit{Low Temp. Phys.} \href{https://doi.org/10.1063/10.0019702}{\textbf{49}, 886--892 (2023)}.
\end{abstract}


\maketitle

\section{Introduction}
One of the fundamental requirements a normal-superconductor (N/S) point contact has to fulfill, in order to allow spectroscopic measurements of the superconducting energy gap, is that the bias voltage $V\ped{c}$ applied to the junction directly measures the excess energy with which the electrons from one bank are injected in the other\,\cite{Naidyuklibro,DagheroSUST2010}. This requires that: i) electrons acquire an excess energy $eV\ped{c}$ while crossing the contact region; ii) the measured voltage $V\ped{exp}$ actually coincides with (or is as close as possible to) the voltage drop at the interface, $V\ped{c}$. 

The first requirements is equivalent to asking that electrons do not undergo inelastic scattering while crossing the contact region; this in turn means that the contact should be in the ballistic regime (i.e. the mean free path is much larger than the contact diameter, i.e. $\ell \gg a$), or, at most, in the intermediate (so-called, diffusive\,\cite{Naidyuklibro}) regime in which electrons can undergo elastic scattering events in the contact region, but not inelastic ones.

The second requirement can never be strictly fulfilled since, as pointed out by Chen et al.\,\cite{Chen2010PRB}, the experimental voltage drop $V\ped{exp}$ is generally measured in a pseudo-four-probe arrangement, and therefore at the ends of a series of resistances: that of the contact and those of the two banks of the junction. For example, if one uses a metallic tip pressed against a superconductor, one has to pay attention to the fact that the voltage drop experimentally measured contains contributions not only from the contact, but also from the tip ($R_1$) and potentially from the superconducting bank ($R_2$) if, for some reasons, it is driven into the resistive state. The overall resistance in series to that of the contact is generally called \emph{spreading resistance}.
The resistance of the tip (if one uses a metal like Ag, Au or Pt) is usually  much smaller than the resistance of the contact itself, and therefore $R_1$ can be disregarded. The resistance of the superconducting bank, $R_2$, can instead play an important role (especially in thin films or two-dimensional (2D) materials) and comes into play when the current drives the superconductor to the normal state. Since the critical current decreases on increasing the temperature, this usually does not affect the conductance spectra at low temperature, but can bend their high-voltage tails and finally determine their overall downward shift when the temperature approaches $T_c$\,\cite{PecchioPRB2013, Daghero2014, Doring2014JPCS}. This effect occurs independently of whether the point contact is ballistic or not, because it involves the bulk of the superconductor. A correction of the experimental conductance curves aimed at eliminating this unwanted effect (that heavily affects the normalization of the spectra) was proposed by Paul Seidel's group\,\cite{Doring2014JPCS} who studied the effect of the spreading resistance on the conductance spectra of planar hybrid SNS' junctions. The junctions were  made by using a thin film of  Ba(Fe,Co)$_2$As$_2$ as base electrode, separated by a gold barrier layer from a Pb counterelectrode\,\cite{Doring2012PhysicaC}. Thanks to the geometry of the system, the authors were able to characterize the electrodes and the junction separately, by using eight electrical connections. They thus directly measured the differential resistance of the superconducting film, $R_{2}=\mathrm{d}V/\mathrm{d}I$, as a function of the current $I$, at any temperature. Then, they showed that subtracting the contribution of the current-dependent spreading resistance from the conductance spectra {\color{blue}{allowed correcting the latter in such a way that the anomalous bending and shift disappeared}}\,\cite{Doring2014JPCS}.

\begin{figure*}[t!]
    \centering
    \includegraphics[width=0.8\linewidth]{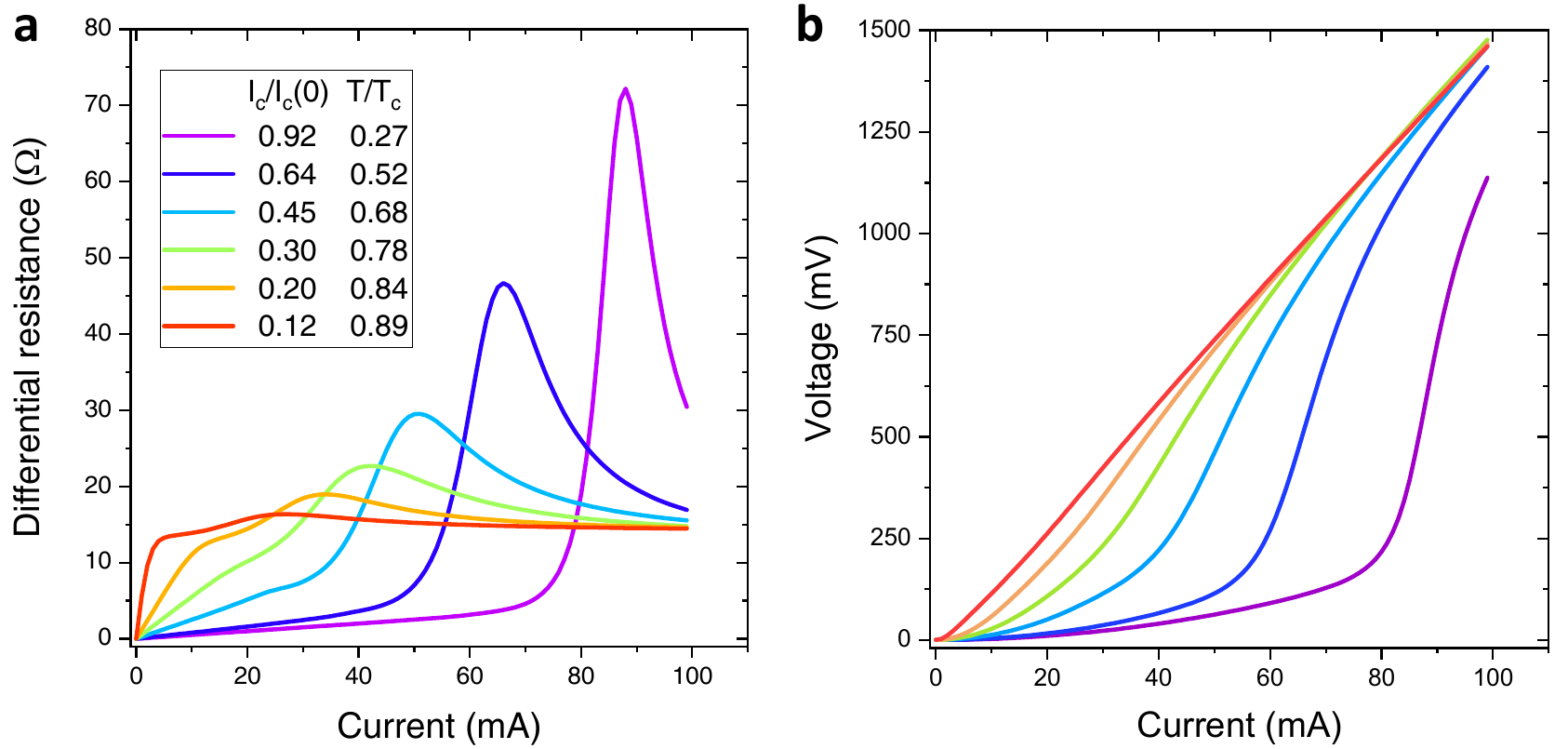}
    \caption{
    (a) Differential resistance d$V$/d$I$ as a function of the current intensity $I$ for different values of $I_c/I_c(0)$, generated by our model in order to fit the experimental curves of Ref.\,\onlinecite{Doring2014JPCS}. Here, $I_c(0) = 85$ mA. The correspondence between $I_c/I_c(0)$ and $T/T_c$ is constructed by assuming a typical behaviour of the critical current as a function of temperature [Eq.\,\eqref{eq:Jc_vs_T}]. (b) The corresponding $V(I)$ characteristics, obtained by integration. The critical current $I_c$ that appears in the model corresponds to the current at which the $V-I$ curves crossover from the almost linear TAFF regime to the typical superlinear flux creep regime\,\cite{Sudbo2004}.
    }
    \label{fig:res_vs_current}
\end{figure*}

Here we will focus on the so-called "dips" that are often seen in point-contact spectra on various materials at much lower voltages, showing that they are based on the same physics. In Ref.\,\onlinecite{StrijkersPRB2001} these dips were instead associated to the depression of the superconducting order parameter at the N/S interface. This would lead to Andreev reflection being sensitive to the (smaller) proximity gap, and to quasiparticle transmission being sensitive to the larger bulk gap. Although suggestive, this model cannot explain the occurrence of dips at energies larger than the bulk gap. Our interpretation is in line with that of  Ref.\,\onlinecite{Sheet2004PRB}, where the dips were explained as being due to the point contact not being in the ballistic regime. 
Excluding the case of a purely Maxwell regime, in which no spectroscopy is possible, it is indeed rather common to obtain point contacts whose resistance can be described by means of the Wexler formula\,\cite{Wexler1966PPS, Nikolic1999PRB}
\begin{equation}
R=\frac{2h}{e^2 a^2 k\ped{F,min}^2 \tau}+
\gamma\left (\frac{\ell}{a}\right )\frac{\rho_1+\rho_2}{4a}.\label{eq:R_diff_hetero}
\end{equation}
as if it were a series of two contributions. In this equation, which holds for heterocontacts, the first term is the Sharvin resistance $R\ped{S}$, where $h$ is the Planck constant, $k\ped{F,min}=\min[k\ped{F,1},k\ped{F,2}]$, $a$ is the contact radius, and $\tau$ is a function of the Fermi velocities $v\ped{F,1}$ and $v\ped{F,2}$. The second term accounts for the Maxwell contribution to the contact resistance, $R\ped{M}$, and contains the resistivities of both the banks of the junction. The prefactor $\gamma(\ell/a)$ is a slowly varying function of the Knudsen ratio, that we will approximate to unity.
As pointed out in Ref.\,\onlinecite{Chen2010PRB}, this way of writing the Maxwell resistance is not completely accurate since it intrinsically contains contributions from both the region of the contact and regions of the material far from the contact. As a first approximation, we will however disregard this detail.
Eq.\,\eqref{eq:R_diff_hetero} properly accounts for the prevalence of the Sharvin term ($\propto a^{-2}$) or of the Maxwell one ($\propto a^{-1}$) depending on the values of the contact size $a$. 
As long as the superconducting bank is in the zero-resistance state, $\rho_2=0$ and only a contribution containing the (usually negligible) resistivity of the metallic counterelectrode is present. However, as noted in Ref.\,\onlinecite{Sheet2004PRB}, the contribution of $\rho_2$ appears as soon as the superconductor is driven to the resistive state in the region of the contact by the current flowing through it. Owing to the characteristic shape of the $V-I$ curve of a superconductor, the resistivity of the material is zero for $I =0$, shows a peak at a current close to $I_c$ and then decreases smoothly to a value different from zero. In Ref.\,\onlinecite{Sheet2004PRB}, Sheet et al. were able to show that, by assuming a "model" $V-I$ curve for the superconductor, the current dependence of the Maxwell term gives rise to typical "dips" in the differential conductance of the point contact, in addition to the structures associated to Andreev reflection at the N/S interface. The fit of the Andreev-reflection spectra (including the dips) with a 2D BTK model\,\cite{BlonderPRB1982, KashiwayaPRB1996, Kashiwaya2000RoPP} (that is actually designed for ballistic contacts) was shown to give rise to an overestimation of the gap amplitudes.

In this paper, we will use a similar approach, but we will go further. As a matter of fact, we will develop an analytical, phenomenological model to reproduce the realistic shape of the $V-I$ curve of the superconductor, and show that, by properly inserting the corresponding d$I$/d$V$ in the expression of the point-contact differential conductance, it is possible not only to fit experimental Andreev-reflection spectra that present the dips, but also to recover the proper values of the gaps. This approach has the advantage that, by adjusting the parameters that control the critical current and the shape of the Maxwell resistance as a function of temperature, it is possible to fit the spectra at various temperatures and in different materials.

\section{The model for the Maxwell resistance as a function of current}
In Ref.\,\onlinecite{Doring2014JPCS}, the differential resistance of a Ba(Fe,Co)$_2$As$_2$ thin film was directly measured as a function of temperature. We found a functional form which is able to mimic rather well the shape of these experimental d$V(I)$/d$I$ curves:

\begin{align}
R\ped{s}(I)  = \frac{\mathrm{d}V}{\mathrm{d}I} = &\left[A \left[ \left(\frac{I_c}{I}\right)^{2n}-\left(\frac{I_c}{I}\right)^{n}\right]+B\right]^{-1} + \nonumber \\[2pt]
& + C \bigintss_{\,\,0}^{I} \displaystyle{\frac{1}{1+ e^{- k \left(1-\frac{I}{I\ped{cut}}\right)}} \mathrm{d} I } \label{eq:R_vs_I}
\end{align}
where $A$, $B$, $n$, $C$, $k$ and $I\ped{cut}$ are parameters that control different details of the shape of the curve, and $I_c$ is the critical current. For the temperature dependence of the critical current $I_c$, we used the expression 
\begin{equation}
    I_c (T) = I_c (0) \left[ 1- \left(\frac{T}{T_c}\right)^2\right] \left[1 - \left(\frac{T}{T_c}\right)^4\right]^{1/2}
    \label{eq:Jc_vs_T}
\end{equation}
taken from Ref.\,\onlinecite{Poole_book}, where $I_c(0)$ is the critical current at zero temperature.

In order to explain the role of the various parameters, it is worth analysing some theoretical curves generated by using this model. Fig.\,\ref{fig:res_vs_current}a shows some curves, all having the form of Eq.\,\eqref{eq:R_vs_I}, that fit almost perfectly the experimental curves reported in Ref.\,\onlinecite{Doring2014JPCS}; the relevant $V(I)$ curves, obtained by integration, are instead shown in Fig.\,\ref{fig:res_vs_current}b. The functional form of Eq.\,\eqref{eq:R_vs_I} is clearly able, with a suitable tuning of the parameters, to reproduce the temperature dependence of the spreading resistance as a function of the current.
In particular, the first term in Eq.\,\eqref{eq:R_vs_I} reproduces the peak and the following decrease at high currents of the differential resistance. The second term, proportional to $C$, is the integral of a sigmoid and just provides a linear term at low current, that saturates to a constant value $C I\ped{cut}$ at $I>I\ped{cut}$ (in all curves of the figure, $I\ped{cut} \simeq 0.7 I_c$).
This linear term is more and more important when the temperature is increased, and allows reproducing the fast increase in $R\ped{s}$ at currents smaller than $I_c$ which is observed at high temperature. The parameter $I_c$ is the critical current, here defined as the current that, in the $V(I) $ curves (Fig.\,\ref{eq:R_diff_hetero}b)  marks the transition from the thermally-assisted flux flow (TAFF) regime  to the flux creep regime\,\cite{Sudbo2004}. This corresponds to the departure of the curves of Fig.\,\ref{fig:res_vs_current}b from the low-current almost linear behaviour. 
The width of the peak in the differential resistance curves of Fig.\,\ref{eq:R_diff_hetero}a is controlled by the parameter $n$, while its height is essentially determined by $A$. All the curves tend to a finite and constant value of resistance (that we will call $R_\infty$ in the following) which is related to the value of the parameters $B$ and $C$. If $C=0$, $B = R_\infty^{-1}$; when $C \neq 0$, the additional linear term gives rise to a constant contribution to $R_\infty$ equal to $C\cdot I\ped{cut}$. As for the parameter $k$, it was always about 6.2 in all the curves of Fig.\,\ref{fig:res_vs_current}.

\section{Fit of the experimental spectra}
\begin{figure}
    \centering
    \includegraphics[width=\columnwidth]{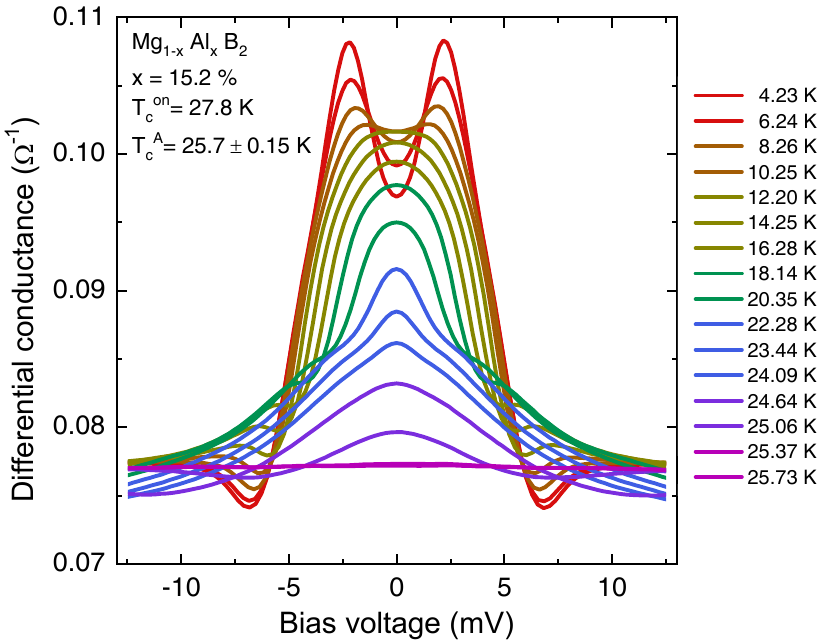}
    \caption{
    Experimental unnormalized spectra measured in a point contact on a Mg$_{0.85}$Al$_{0.15}$B$_2$ single crystal at different temperatures. The onset of the resistive transition of the cyrstal was $T_c^{on} = 27.8$\,K while the Andreev-reflection signal disappears at $T_c^A = 25.5 \pm 0.15$\,K.
    }
    \label{fig:rawcurves}
\end{figure}

Once understood that the functional form of Eq.\,\eqref{eq:R_vs_I} is sufficiently flexible to reproduce the shape of the $V(I)$ and of the $R\ped{s}(I)$ curve experimentally measured in a given material, we tried to apply it to the fit of the experimental d$I$/d$V$ spectra, shown in Fig.\,\ref{fig:rawcurves}, that were measured in a Mg$_{0.85}$Al$_{0.15}$B$_2$ single crystal, and that present very clear dips. 
Note that: i) the material is completely different from the iron-based compound in which the $R\ped{s}(I)$ curves we used to construct the model were measured; ii) our sample is a single crystal and not a film; iii) we are going to use the model of Eq.\,\eqref{eq:R_vs_I} to mimic the current dependence of the Maxwell term $R\ped{M}$ in the contact (and not the onset of the spreading resistance $R_2$ associated to the resistive state of the film, as in Ref.\,\onlinecite{Doring2014JPCS}). Despite the model being used in completely different conditions from those which led to its development, we will show that it provides excellent results.

The spectra of Fig.\,\ref{fig:rawcurves} were {\color{blue} measured in a single crystal of Mg$_{1-x}$Al$_{x}$B$_2$ with $x=0.15$, grown by using the high-pressure cubic anvil technique described in Ref. \onlinecite{Karpinski2005PRB} and by optimizing time, pressure and temperature to avoid any phase segregation up to $x=0.32$. Indeed, no impurities, twins or
intergrowing crystals were detected \cite{Daghero2008JPCM}. Due to the difficulty in obtaining significant Al doping levels in MgB\ped{2}, some small inhomogeneity of the doping content can be expected.
The high quality of the crystals allowed us to obtain spectroscopic contacts and textbooklike PCARS spectra, from which we could extract the energy gaps as a function of the doping content \cite{Daghero2008JPCM}. The curves shown in Fig.\,\ref{fig:rawcurves} are instead ``non-ideal'' spectra that were already shown, although with a vertical offset, in Ref.\,\onlinecite{DagheroSUST2010} as a perfect example of the temperature dependence of the dips.} Here, the absence of any offset allows appreciating the position of the normal-state conductance curve with respect to the superconducting ones. The Andreev-reflection features completely disappear somewhere between 25.37 and 25.73\,K and we thus defined $T_c^A = 25.5 \pm 0.15$\,K. By the way, the onset of the superconducting transition in this crystal was at $T_c^{on} = 27.8$\,K: {\color{blue}this difference may be ascribed to the aforementioned inhomogeneity in the Al content, but also to a small heating in the contact\,\cite{Naidyuklibro, Verkin79}, which is expected to take place as soon as the Maxwell term appears.}
It is clear from Fig.\,\ref{fig:rawcurves} that, at low temperature, the high-energy tails of the experimental curves still tend (from above) to the experimental normal-state conductance, despite the presence of clear dips between 5.0 and 7.5\,mV. However, above 20\,K, the high-energy tails lie \emph{below} the normal state at $T_c$. This is clearly due to the breakdown of superconductivity in the bulk, i.e. the effect described and treated by D\"oring et al.\,\cite{Doring2014JPCS} that we are not considering here. Therefore, if we want to concentrate on the curves that \emph{only} present dips due to the Maxwell term, we have to focus on the range of temperatures between 4.2 K and 18.14\,K. 

Let us thus use Eq.\,\eqref{eq:R_vs_I} to model the current dependence of the Maxwell resistance $R\ped{M}$ that, according to Eq.\,\eqref{eq:R_diff_hetero}, can be considered to be in series with the contact resistance. Therefore, we can compute the $R\ped{M}(I)$ curve, that clearly depends on the parameters listed above. 

Then, we calculate the theoretical normalized conductance associated to Andreev reflection in the contact, $\sigma\ped{BTK}$, according to the 2D BTK model\,\cite{BlonderPRB1982, KashiwayaPRB1996, Kashiwaya2000RoPP} generalized to the case of two gaps\,\cite{DagheroSUST2010, Daghero2011} since the material under study is Al-doped MgB$_2$\,\cite{Karpinski2005PRB, Daghero2008JPCM}.
The parameters required to calculate $\sigma\ped{BTK}$ are the gap amplitudes $\Delta_1$ and $\Delta_2$, the barrier parameters $Z_1$ and $Z_2$, the broadening parameters $\Gamma_1$ and $\Gamma_2$ and the relative weight of the contribution of the first gap to the spectra, $w_1$ (such that $w_2= 1- w_1$). 
The calculated theoretical \emph{normalized} curve, $\sigma\ped{BTK}$, must then be multiplied by the normal-state conductance in order to get the unnormalized differential conductance. However, one cannot use for this purpose the \emph{actual} normal-state conductance $G\ped{n}\apex{\,exp}$ because the appearance of an additional resistance in series with the Sharvin one [when $\rho_2$ starts to be different from zero, see Eq.\,\eqref{eq:R_diff_hetero}] not only makes the measured resistance of the whole series increase (thus shifting the normal-state conductance curve downwards\,\cite{PecchioPRB2013, Daghero2014}) but also makes the experimental voltage $V\ped{exp}$ be different from the voltage drop across the contact\,\cite{Chen2010PRB, Sheet2004PRB, Doring2014JPCS}, $V_c$, because
\begin{equation}
    V\ped{exp}(I) = V\ped{c}(I) + \int_0^I R_M(I') dI' \label{eq:Vcorr}
\end{equation}
The stretching of the voltage scale, together with the downward shift due to the additional $R\ped{M}$ term, implies that the experimental normal-state conductance $G\ped{n}\apex{\,exp}$ is smaller, and extended to higher voltages, than the hypothetical normal-state conductance curve that one would measure if the contact was ballistic, i.e. $G\ped{n}\apex{\,ideal}$. 
However, the "ideal" normal-state conductance can be reconstructed by inverting the experimental one, subtracting $R_{\infty}$, and correcting the voltage scale according to Eq.\,\eqref{eq:Vcorr}.

The unnormalized BTK conductance of the junction is thus
\begin{equation}
    \frac{\mathrm{d}I}{\mathrm{d}V\ped{c}}(V\ped{c}) = \sigma\ped{BTK}(V\ped{c}) \, G\ped{n}\apex{\,ideal} (V\ped{c}). 
\end{equation}
This, once inverted (to get the unnormalized BTK resistance) and expressed as a function of the current, can be summed to the Maxwell term $R\ped{M}(I)$, thus providing the total resistance of the series. This result must be inverted again, giving the total conductance (including the dips). At the end of the process, one can express the total differential conductance as a function of the total voltage. This curve can be directly compared to the experimental unnormalized conductance as a function of $V\ped{exp}$. It is thus possible to find the best set of parameters that makes the theoretical curve properly fit the experimental one.

\begin{figure}
    \centering
    \includegraphics[width=0.8\linewidth]{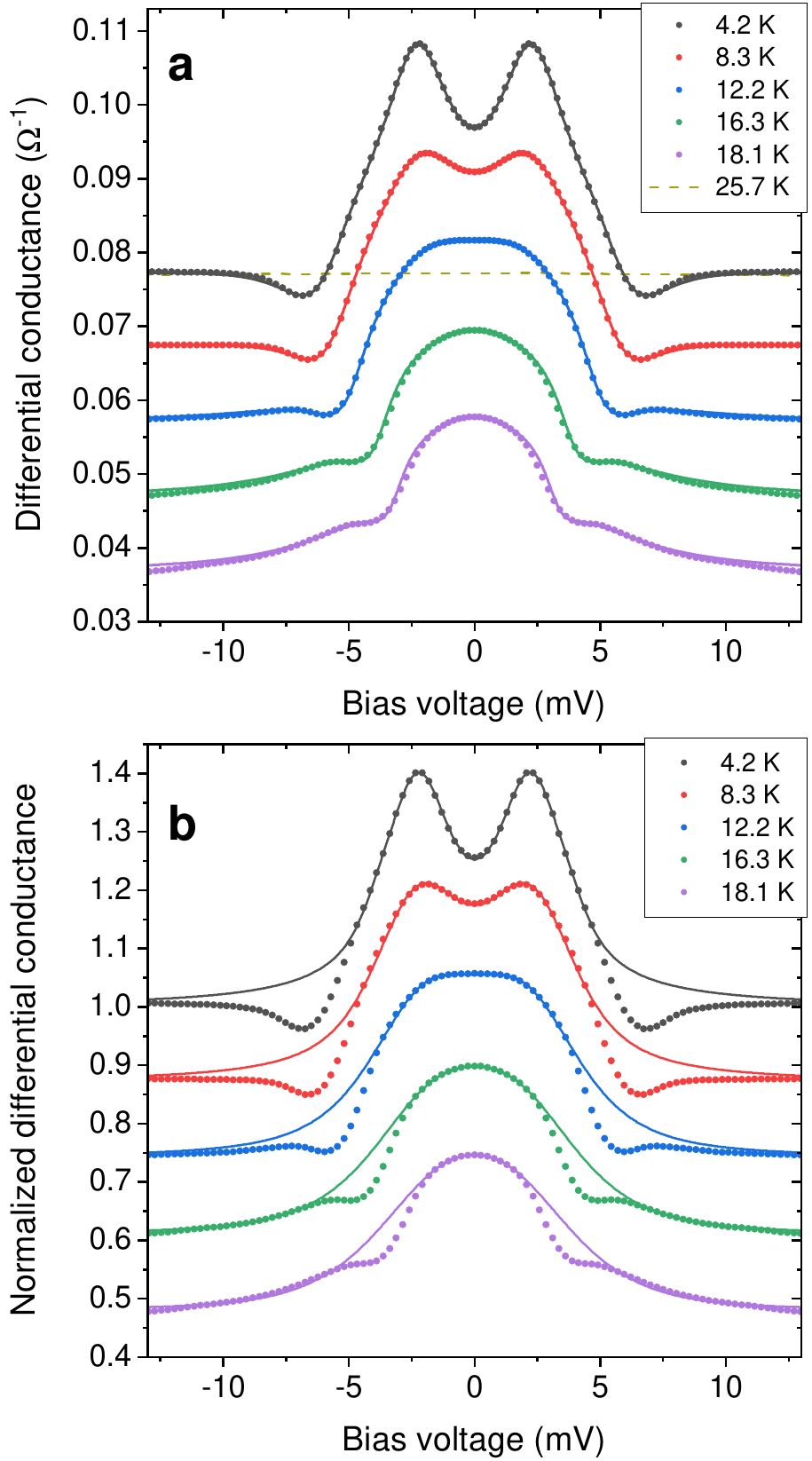}
    \caption{
    (a) As-measured, unnormalized differential conductance d$I$/d$V$ of a soft point contact on Mg$_{0.85}$Al$_{0.15}$B$_2$ single crystal, at different temperatures (symbols) together with the relevant fits (solid lines) that include both Andreev reflection and the current-dependent Maxwell term. The dashed horizontal line is the normal-state conductance at $T=25.7$\,K. All the curves have been vertically offset for clarity, apart from the lowest-temperature one (black symbol) and the normal-state one. (b) Experimental point-contact curves, normalized to the actual normal-state conductance at $T = 25.7$\,K and fitted to the standard 2D BTK model.
    }
    \label{fig:fits}
\end{figure}

Fig.\,\ref{fig:fits}a reports the results of the fit of the raw conductance curves of Fig.\,\ref{fig:rawcurves}, up to 18.1\,K. The symbols represent the experimental curves and the lines the fitting functions. For the sake of comparison, panel (b) of the same figure displays the fit one would obtain by using the two-band, 2D BTK model, without accounting for the dips. 
The quality of the fit in Fig.\,\ref{fig:fits}a is extremely good despite the fact that the functional form of the Maxwell term [Eq.\,\eqref{eq:R_vs_I}] was taken from the experimental $R\ped{s}(I)$ curves of a film of a completely different material. This suggests that the functional form of Eq. \ref{eq:R_vs_I} is very general, and can be adapted (by suitably choosing the parameters) to different cases.  On the contrary, the fit shown in Fig.\,\ref{fig:fits}b is reliable at low temperature, when the dips fall at an energy slightly larger than the large gap, but becomes more and more meaningless on increasing the temperature, since the dips shift to lower voltages and end up by heavily interfering with the gap structures.

\begin{figure}
    \centering
    \includegraphics[width=0.9\columnwidth]{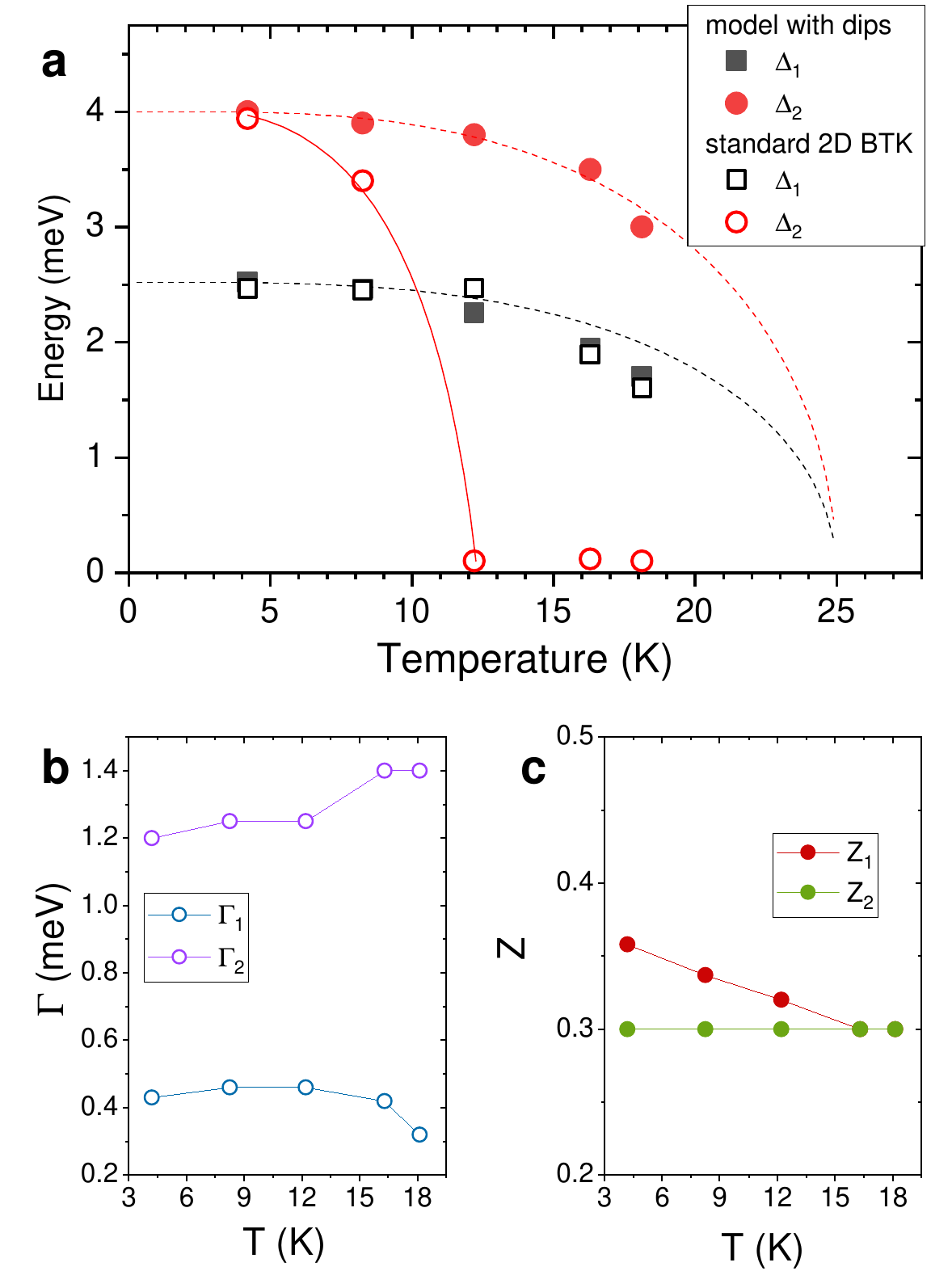}
    \caption{
    (a) Temperature dependence of the energy gaps used to fit the experimental conductance curves. Solid symbols indicate the gaps obtained by using the 2D BTK model including dips (as in Fig.\,\ref{fig:fits}a), while open symbols represent the gaps obtained by using the 2D BTK model alone (as in Fig.\,\ref{fig:fits}b).  The dashed lines indicate the BCS-like temperature dependencies. (b) Temperature dependence of the broadening parameters $\Gamma_1$ and $\Gamma_2$. (c) Temperature dependence of the barrier parameters $Z_1$ and $Z_2$. Data shown in (b) and (c) are obtained by using the 2D BTK model including dips and the lines are just guides to the eye.
    }
    \label{fig:parametersBTK}
\end{figure}

The values of the gap amplitudes extracted from the fit of the experimental conductance curves are shown in Fig.\,\ref{fig:parametersBTK}a. The model that includes dips gives the gap amplitudes $\Delta_1$ (the smaller) and $\Delta_2$ (the larger), indicated by solid symbols, which follow rather well a BCS-like trend (dashed lines). The fit with the 2D BTK model alone, instead, provides the same values of the gap amplitudes only at 4.2\,K. On increasing the temperature, the large gap $\Delta_2$ (red open circles) immediately decreases, because of the shift of the dips to lower energies; the uncertainty on the amplitude of $\Delta_2$ becomes larger than $\Delta_2$ itself already at 8.3\,K, and at higher temperatures the large gap is completely undetermined (the fit converges to a small value of $\Delta_2$, but with a huge uncertainty, meaning that the fit is possible as well with a single gap). The values of the small gap (black open squares) do not deviate very much from those provided by the model with dips, just because $\Delta_1$ is much less affected by the presence and the displacement of the dips. 

Fig.\,\ref{fig:parametersBTK}b and c report the temperature dependence of the other fitting parameters contained in the 2D BTK part of our model. The broadening parameter $\Gamma_1$, associated with the small gap, remains practically constant around 0.4\,meV, and is thus much smaller than the gap amplitude $\Delta_1$; it just had to be slightly decreased at the highest temperature. $\Gamma_2$, associated with the large gap, is much smaller than $\Delta_2$ in the whole temperature range, and increases from 1.2\,meV at 4.2\,K to 1.4\,meV at 18.1\,K. The barrier parameters, in principle, should not change with temperature, being related to the potential barrier at the interface and to the mismatch of Fermi velocities on the two banks\,\cite{DagheroSUST2010, Naidyuklibro}. Indeed, $Z_2$ is perfectly constant and equal to 0.3; $Z_1$ instead had to be slightly reduced on increasing temperature to perfectly fit the low-bias region of the curves (the maximum variation is however $< 16\%$). By the way, the same fitting parameters acquire non-physical values in the case of the pure 2D BTK model, further indicating its inadequacy. In particular, $\Gamma_2$ is almost zero at low temperature and jumps to 5\,meV already at 12.2\,K, while $\Gamma_1$ is about 0.5\,meV at low temperature and decreases to zero on increasing the temperature. In either case (i.e. 2D BTK model, or model with dips) we kept the weight $w_1$ constant at all temperatures. In particular, $w_1 =0.724$ in the fit with dips, and $w_1 =0.892$ in the 2D BTK fit. 

\begin{figure}
    \centering
    \includegraphics[width=0.9\columnwidth]{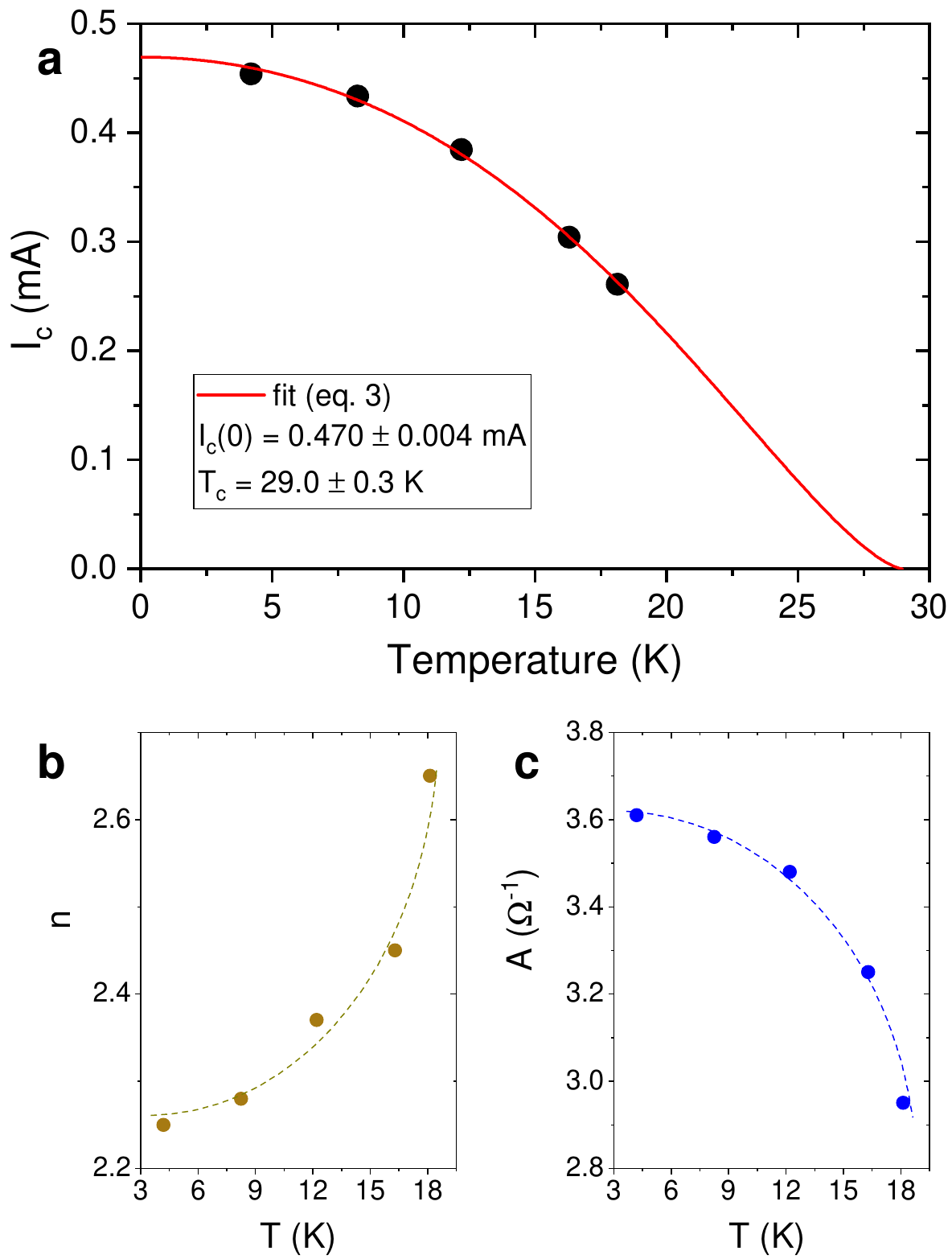}
    \caption{
    (a) Temperature dependence of the critical current \emph{in the contact region} as extracted from the fit of the conductance curves (symbols) and the relevant fit with the theoretical $I_c(T)$ curve [Eq.\,\eqref{eq:Jc_vs_T}]. The fitting is almost perfect, with $I_c (0) =0.470 \pm 0.004$\,mA, and $T_c = 29.0 \pm 0.3$\,K. (b) Temperature dependence of the parameter $n$ that enters Eq.\,\eqref{eq:R_vs_I}. (c) Temperature dependence of the parameter $A$ that enters Eq.\,\eqref{eq:R_vs_I}. Dashed lines in (b) and (c) are just guides to the eye.
    }
    \label{fig:parametersMaxwell}
\end{figure}

As for the parameters that control the shape of the Maxwell resistance as a function of temperature, we fixed the values of some of them in order to reduce as much as possible their number. 
First of all, with reference to Eq.\,\eqref{eq:R_vs_I}, we kept $C=0$. This means that, unlike in the case shown in Fig.\,\ref{fig:res_vs_current}, the fit did not require any low-current linear behaviour of the $R(I)$ curves. The fact that $C=0$ also implies that the parameters $k$ and $I\ped{cut}$ do not appear any longer, and that $B = R_\infty^{-1}$. This is fixed by the vertical shift of the conductance in the normal state, due to the additional Maxwell term. This value was kept constant as a function of temperature, i.e. $R_\infty = 0.26 \,\Omega$ that implies $B=3.846 \,\Omega^{-1}$. Therefore, the only parameters that we changed as a function of temperature are $A$, $n$ and $I_c$, and their behaviour is shown in Fig.\,\ref{fig:parametersMaxwell}. In particular, $I_c$ decreases monotonically as a function of temperature (solid symbols in Fig.\,\ref{fig:parametersMaxwell}a) with a trend that can be very well fitted by the temperature dependence of Eq.\,\eqref{eq:Jc_vs_T}. The fit of the data points with that function provides $I_c(T=0)= 0.470 \pm 0.004$\,mA, and $T_c=29.0 \pm 0.3$\,K. The critical temperature is higher than the experimental $T_c^A$, and also than the temperature at which the superconducting transition starts ($T_c^{on}=27.8$\,K).  {\color{blue}{This mismatch can be partly due to the heating effect in the contact\,\cite{Naidyuklibro, Verkin79} but it may also arise from the fact that}}, MgB$_2$ being a two-band superconductor, Eq.\,\eqref{eq:Jc_vs_T} may not perfectly reflect the temperature dependence of the critical current\,\cite{Nicol2005PRB}. Note that the values of $I_c$ here obtained represent the values of the current \emph{in the contact} that makes the resistivity of the material become different from zero (because of vortex motion).
As for the other parameters of Eq.\,\eqref{eq:R_vs_I}, Fig.\,\ref{fig:parametersMaxwell}b and c show that $n$ increases by 17\% on increasing the temperature, i.e. from 2.25 to 2.65, while $A$ decreases as a function of $T$, from 3.61\,$\Omega^{-1}$ to 2.95\,$\Omega^{-1}$ (thus changing by about 18\%).
This behaviour suggests some interplay between $A$ and $n$ and, indeed, their product is almost constant, ranging from 8.12\,$\Omega^{-1}$ at 4.2\,K to 7.82\,$\Omega^{-1}$ at 18.1\,K. 

Overall, accounting for the dips required three parameters in addition to those included in the two-band, 2D BTK model. However, these parameters control the shape and the position of the dips and, once adjusted so as to obtain a very good fit to the experimental curve, allow obtaining the value of the large gap $\Delta_2$ even though the structures associated to this gap are visibly eroded by the dips. Note that the case-study we have chosen here is actually a particularly critical one since, even at the lowest temperature, the dips are very close to the large-gap structures. In many cases, fortunately, the dips lie far apart from the gaps at low temperature and start to interfere with them only at higher temperatures. In these cases, the use of this model can allow determining the gaps in a wider temperature range than the standard 2D BTK model.

\section{Conclusions}
It is rather common, especially in superconductors with small mean free path, to obtain Andreev-reflection spectra that display, in addition to the gap structures, typical dips that have been associated to the onset of a Maxwell contribution to the contact resistance. This, in turn, occurs when the injected current makes the material become resistive. The presence of dips heavily complicates the process of normalization of the spectra, and may prevent their fit with standard models for Andreev reflection, such as the BTK\,\cite{BlonderPRB1982} model or its generalizations \cite{KashiwayaPRB1996,Kashiwaya2000RoPP}.
Here, we found a phenomenological functional form for the resistance of a superconductor as a function of current that is able to reproduce experimental measurements as a function temperature \cite{Doring2014JPCS}. Then, we showed that the inclusion of this term in the expression of the differential conductance of a point contact, together with the model for Andreev reflection, allows a very good fit of the experimental spectra without requiring their normalization. The fit provides a corrected value of the energy gaps and also allows obtaining the temperature dependence of the critical current intensity in the contact. 


\section*{Acknowledgments}
We are thankful to F. Laviano and S. Sparacio for fruitful scientific discussions.

\bigskip
%

\end{document}